# Prolonging Network Lifetime in Wireless Sensor Networks with Path-Constrained Mobile Sink

Basilis G. Mamalis

Department of Informatics
Technological Educational Institute of Athens
Athens, Greece

*Abstract*—**Many studies in recent years have considered the use of mobile sinks (MS) for data gathering in wireless sensor networks (WSN), so as to reduce the need for data forwarding among the sensor nodes (SN) and thereby prolong the network lifetime. Moreover, in practice, often the MS tour length has to be kept below a threshold, usually due to timeliness constraints on the sensors data (delay-critical applications). This paper presents a modified clustering and data forwarding protocol combined with a MS solution for efficient data gathering in wireless sensor networks (WSNs) with delay constraints. The adopted cluster formation method is based in the 'residual energy' of the SNs and it is appropriately modified in order to fit properly to the requirement of length-constrained MS tour, which involves, among else, the need for inter-cluster communication and increased data forwarding. In addition, a suitable data gathering protocol is designed, based on an approximated TSP route that satisfies the given length constraint, whereas the proper application of reclustering phases guarantees the effective handling of the 'energy holes' caused around the CHs involved in the MS route. Extended simulation experiments show the stable and energy-efficient behavior of the proposed scheme (thus leading to increased network lifetime) as well as its higher performance in comparison to other competent approaches from the literature.**

*Keywords—wireless sensor; mobile sink; node clustering; data gathering; network lifetime*

## I. INTRODUCTION

The interest in the use of WSNs has grown enormously during the last decade, pointing out the crucial need for efficient and reliable routing and data gathering protocols in corresponding application environments. Energy efficiency is one of the main design goals in a WSN, towards the above direction. Moreover, the appropriate minimization of nodes energy consumption as well as the uniform energy depletion of all nodes, are critical parameters in order to increase the time the network is fully operational. In typical WSNs a main reason of energy depletion concerns the need for transmitting the sensed data from the sensor nodes (SNs) to remote sinks.

These data are typically relayed using ad hoc multi-hop routes in the WSN. A side-effect of this approach is that the SNs located closer to the sink are heavily used to relay data from all network nodes; hence, their energy is consumed faster, leading to a non-uniform depletion of energy in the WSN [1]. This results in network disconnections and limited network lifetime. Several protocols have been proposed so far for efficient data gathering in WSNs taking also into account the above problem in order to increase the lifetime of the WSN. The most promising of them involve the mobility of the sink, based on the key idea of changing progressively the neighbors of the sink so that the energy consumption for data relaying is balanced throughout the network [1]. The MS may visit each SN and gather its data [2-3] (single-hop communication) or may visit only some locations and the SNs send their data to the MS through multi-hop communication [4-9]. The delay in data gathering is minimized appropriately in the latter case, however special attention has to be given in the increased energy consumption due to the multi-hop communication used for data forwarding.

A solution in between is to have the SNs send first their data to a certain number of intermediate nodes (building direct or indirect hierarchical clustering structures) which buffer the received data and send them to the MS when it comes within their transmission range or when they receive a query from the MS asking for the buffered data [10-18]. Most of these approaches naturally strike the balance between the data gathering delay and the energy consumption overhead, whereas also, they are usually highly effective in applications where there are restrictions with regard to the sites that can be visited by the MS. Some of these works (like the one presented in [10]) have also faced effectively the problem of the energy-holes caused around the intermediate data-relaying nodes / CHs. Furthermore, in [19] a relevant and very promising structured approach (a geographic convergecast based method) is proposed aiming to reduce the path reconstruction cost upon sink mobility. In the proposed algorithm a virtual backbone structure is formed which is comprised of several virtual circles and straight lines where the CHs are placed adequately.

The primary disadvantage of most of the above approaches is the increased latency of data collection. Indeed, the typical speed of a MS is quite restricted, thus resulting in substantial travelling time and, correspondingly, delay in gathering the sensors data. In practice, often the MS tour length is bounded by a pre-defined time deadline, usually due to timeliness constraints on the sensors data.

Actually, not many works deal with this problem in the general case (i.e. getting the constraint as a parameter and fixing their solutions accordingly); and most of them adopt the

This research has been co-financed by the European Union (European Social Fund – ESF) and Greek national funds through the Operational Program "Education and Lifelong Learning" of the National Strategic Reference Framework (NSRF)–Research Funding Program: Archimedes III. Investing in knowledge society through the European Social Fund.

approach of using multiple MSs, trying to optimize the total tour time for all paths [18,25-26].

To address the above problem in the general case, some of the most notable proposals [20-21] adopt the hybrid approach which combines multi-hop forwarding with the use of a MS which visits only some locations (caching points - CPs), building direct or indirect hierarchical clustering structures. Especially in [20], the problem is addressed as an optimization problem, and the authors focuses on minimizing the total number of forwarding hops from all SNs to their respective nearest CPs. The heart of the proposed solution is a k-means inspired node-clustering algorithm where the main idea is grouping the network into a number of balanced-size clusters, then constructing the MS tour to involve one CP from each cluster, and then iterating over these two phases until the ideal (maximum) number of clusters is found (subject to the constraint of the MS tour length). The proposed solution is evaluated experimentally on a wide range of practical scenarios, showing that it consistently outperforms the algorithm of [21] and is not very far from the optimum in small instances. The problem of planning multiple MS paths that optimize the total length travelled while gathering the data has also been investigated by the same authors in [22,23].

In our work we propose an alternative solution that is based mainly in the 'residual energy' of the SNs (instead of 'distance' - number of hops - in [20]) and aims to take advantage of the energy stable and efficient behavior offered by hierarchical clustering structures in order to increase the network lifetime. We first use as our base node-clustering algorithm the multi-hop clustering algorithm of [24] (which adopts as the main cluster formation criterion the residual energy of each SN), in order to gain energy-balanced clusters that guarantee almost ideal behavior (in terms of average energy consumption and network lifetime) in the special case that no tour-length constraints are given (i.e. the MS can visit all the CHs locations - see also [10]). We then appropriately modify this algorithm in order to fit properly to the requirement of length-constrained MS tour (which involves, among else, the need for inter-cluster communication and increased data forwarding), and we develop a suitable data gathering protocol based on an approximated TSP route that satisfies the given length constraint. Moreover, in order to face up effectively the 'energy holes' that are naturally caused around the CHs involved in the TSP route, we apply a combined scheme that involves proper reclustering phases along with alternating between different initial locations of the MS.

Furthermore, we have performed extended simulation experiments, which show the high performance of our data gathering scheme (in terms of balanced energy consumption and network lifetime), either in the case of pre-defined delay constraints or in the ideal case of no such constraints. Moreover, the corresponding simulation results have shown that our data gathering scheme has considerably better behavior, according to both the above measures, when compared to the corresponding scheme of [20]; which is one of the most relevant and competent works in the literature.

As also mentioned earlier, a usual alternative towards the same direction is to employ more than one MSs. More concretely, many of the most recent attempts in the literature focus on the appropriate generalization of the ideas used with a single MS, on large and very large WSN environments with the use of multiple MSs ([18,25-26]) or the use of mobile relay nodes (MRNs [27-29]), in order to achieve both low energy consumption and reduced total data gathering delay (when compared to the case of a single MS). However this solution is often impractical due to the relatively high cost of the mobile elements used as well as the additional costs required for management and coordination. A relevant extensive survey on data gathering with mobile elements, giving emphasis in the internals of the data collection process (discovery, data transfer, routing etc.) can be found in [30], whereas a corresponding survey focusing in the protocols and algorithms used can be found in [31,32].

The rest of the paper is organized as follows. In section II, the brief description of our base node-clustering algorithm is given. In section III, the proposed modified clustering and data forwarding scheme is presented. In section IV the complete data gathering protocol is given along with the adopted global reclustering scheme. Section V outlines the experimental results, whereas section VI concludes the paper.

## II. THE INITIAL CLUSTERING ALGORITHM

As mentioned above, our overall data gathering solution is based on clustering, in which some nodes (elected CHs) collect/buffer the received data from other nodes, and send them to the MS when it comes within their transmission range. We use multi-hop clustering in order to be able to control latency by having the MS visit the locations of smaller number of nodes, i.e. the locations of the elected CHs, as well as to properly control, balance and restrict the multi-hop communication overhead.

Moreover, we use as our base clustering algorithm the relevant multi-hop algorithm of [24]. This algorithm adopts as the main cluster formation criterion the 'residual energy' of each SN), and leads to energy-balanced clusters, as well as to effective handling of the 'energy holes' caused around the CHs, thus prolonging the network lifetime. Specifically, the cluster formation algorithm of [24] consists of the following steps:

- Initially, all the nodes in the network broadcast messages (including their residual energy and their ID) in a certain power, which ensures that nodes within a radius R (which is a pre-defined threshold) will receive the message. Then, each node waits to receive such messages from all its 'neighboring' (within a radius R) nodes.

- For every such received message, each node compares the residual energy in the message with its own energy, and then it acts as follows: If the energy in the message is larger, it marks the node which sent the message as its parent.

- If the node which received the message has already a parent, and the node which sent the message has larger residual energy than its own, then it compares the distance between it and its parent with the distance between it and the node which sent the message. If the former is larger, it replaces its parent with the node who sent the message.

- When a node has received all its neighboring nodes messages and has made the necessary decisions, it sends a 'join' message to its parent node, and marks itself as a 'member' node.

- If the node has no parent node, then it marks itself as a CH and broadcasts a relevant message.

Thus, at the end of execution each node has as 'parent' the node that has larger residual energy than its own, with the minimum distance. Experimental results in [24] show that the above algorithm effectively saves the energy costs, leads to balanced energy consumption and prolongs the lifetime of the network. Furthermore, the main goal of the algorithm is the creation of suitable (energy-balanced) clusters with not only high-energy CHs, but also having energy-rich neighborhoods.

In that way it effectively avoids energy holes around the CHs and naturally it becomes quite suitable for energy-efficient data gathering using a MS. Specifically, by following a simple data gathering protocol (e.g. having the MS scheduled to visit all the elected CHs through an appropriately computed optimal distance TSP path and gather the sensed data, like in [10] - see also fig. 1), one should normally expect to achieve a quite efficient total gathering solution (preserving low-variance energy consumption and high levels of network lifetime), suitable for applications with no tour-length constraints.

The simulation experiments presented in section V (fig. 8) validate the above conclusion. We also choose the algorithm of [24] as our base clustering algorithm because it is completely distributed and localized, it spends low number of messages, and fits well to the orientation restrictions applied in the modified version that directly follows.

## III. THE PROPOSED MS ORIENTED CLUSTERING ALGORITHM

Towards the direction of developing a relevant data gathering solution (taking into account the residual energy as the main criterion for data forwarding) that also satisfies specific tour-length constraints, we first proceed to a suitable extension of the node-clustering algorithm of [24], in such a way that its main characteristics still hold and the necessary communication between the elected CHs is efficiently performed. The main goal of the proposed extension is to build an energy-efficient total solution, that will be primarily suitable for delay-critical applications (i.e. applications that the sensed data have to be gathered/uploaded to the Base Station (BS) within a specific - usually periodic - short range of time, e.g. L), especially over large-scale sensors deployment areas.

We assume that all the deployed SNs have the same equipment, they start with uniform energy, each SN knows its location and no aggregation takes place (all the sensed data have to be sent to the MS). Within the above context, the proposed extension consists of two basic rules, one with regard to cluster formation and one specifying how data forwarding between the CHs (inter-cluster communication) should be performed. The detailed description of these basic rules directly follows (in figures 2-4), along with corresponding explanations and discussion.

The main objective of Rule 1 (fig. 2) is to form the final clusters in such a way that unnecessary transmissions from a member-node (back to the CH - at the opposite direction - and then forward from the CH to the MS at the straight direction) during the final data gathering phase, are strictly avoided. In other words, all the member-nodes of a cluster should have their CH in the same 'direction' with the MS; so as all the necessary data forwarding from each member-node (first towards its CH and then from the CH towards the MS) take place in one direction only.

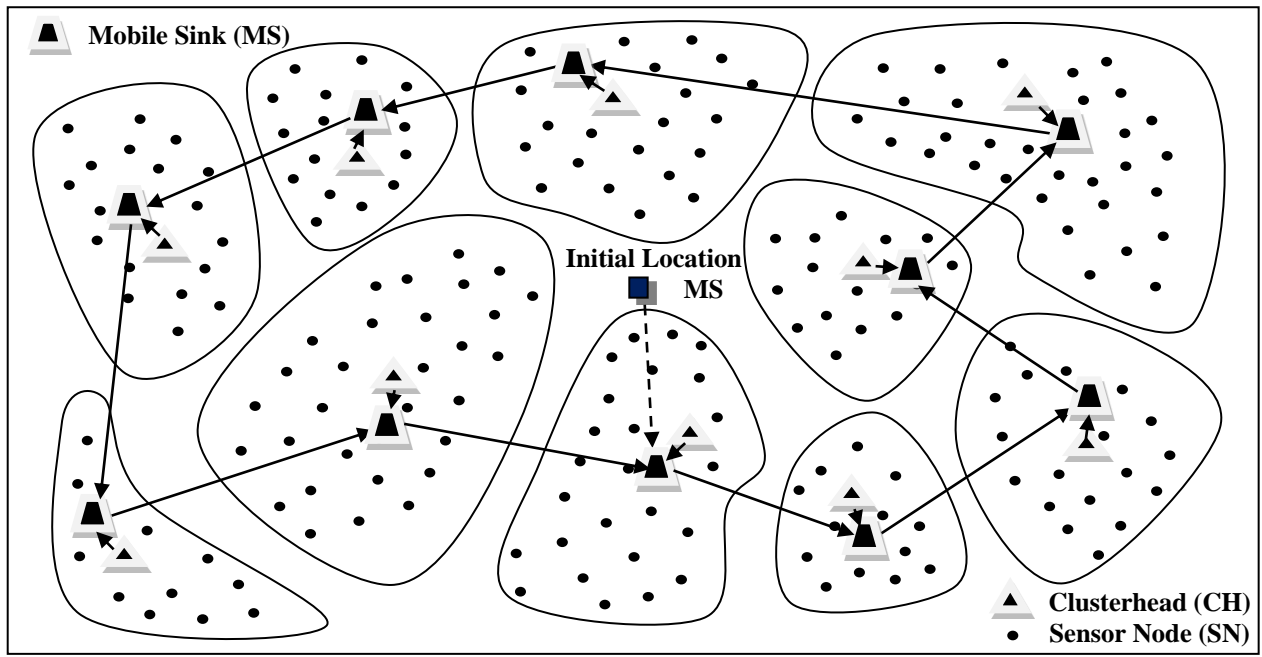

Fig. 1. Data gathering with the initial clustering scheme (without constraints)

**Rule 1:**

Execute the node-clustering algorithm of [24] with the following change: allow each SN to be attached (mark as 'parent') only to neighboring nodes that lie within the 'direction' from the SN to the MS.

This can be simply done by setting each SN in step 1 of the algorithm (before proceeding to steps 2,3 and make the necessary comparisons), take into account only the received messages that come from SNs that satisfy the above orientation restriction. Also, each SN should include its location (along with its residual energy and its ID) in its initial broadcasting message.

The decision if a neighboring node $v$ lies within the 'direction' from the SN to the MS is based on the angle $\theta$ between the lines connecting the SN to $v$ and to the MS respectively, as shown in fig. 3.

Fig. 2.    The 1st rule of the MS-oriented clustering

As it can be easily noticed, the higher the value of $\theta$ the higher the probability for each SN to find enough nodes with larger residual energy (through Rule 1), as well as the broader the area in front of each CH that the total load of forwarded data will be distributed (later on, through Rule 2). However also, the higher the value of $\theta$ the higher the probability of concluding to longer paths for some of the communicated messages (either within each cluster after Rule 1 or between the CHs through Rule 2), thus loosing the advantage gained from the balanced energy consumption that lies in the heart of our clustering algorithm. Generally, the best choice for the value of $\theta$ depends on the actual distribution of the SNs in the deployment area as well as on the density of that distribution. Naturally, an optimal value of $\theta$ is expected to be determined only experimentally. In that sense (as it comes out from our simulation results - section V) a value of $\theta$ between 65° and 75° is likely to lead to the best results in terms of total energy efficiency (average consumption and variance). At the end of the execution, one CH will be elected around the 'top' of each cluster and it will be the root of a suitably balanced-energy node-tree (see fig. 6).

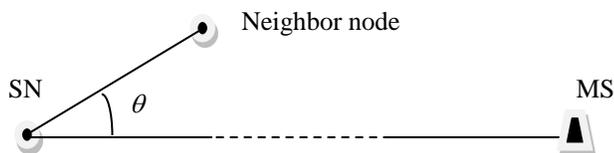

Fig. 3.    Neighbor nodes orientation with respect to MS

Considering the simple case of delay tolerant applications (i.e. without any latency restrictions) no inter-cluster communication (data forwarding between the CHs) is needed due to the fact that the MS is scheduled to visit all the CHs locations. On the contrary, if specific time constraints with respect to the total data gathering delay are to be satisfied (e.g. within the context of a large-scale WSN application), inter-cluster communication is necessary since the MS should be scheduled to visit only a subset of the CHs locations (as it is shown in section IV). Moreover, the way this inter-cluster communication will be performed, is obviously crucial with regard to the total energy efficiency of the whole solution.

**Rule 2:**

Each CH forwards the received data (either the sensed data of the member-nodes of its cluster or the sensed data from nodes of other clusters that are forwarded by the member-nodes of its cluster), in a round-robin manner, to neighboring nodes that belong to other clusters and lie within the 'direction' from the CH to the MS (see fig. 3).

Note that the set and the locations of the candidate neighbors (the ones lying within the direction from the CH to the initial location of the MS) are known from the initial execution of the clustering algorithm (see Rule 1). Let's denote this set as N, the number of these nodes as |N| and their IDs as $N_i$ where i = 0…(|N|-1). Then, typically each CH should forward the received messages in the following manner:

$i = 0;$
*For each* outgoing message $M$ (i.e. message that has to be forwarded to the *MS*)
    Send $M$ to neighbor node $N_{i\%|N|}$ ;
    i = i + 1;

Fig. 4.    The 2nd rule of the MS-oriented clustering

Specifically, the corresponding messages of each CH should be forwarded towards the MS in such a way that the total communication load is distributed evenly to the intermediate (from the specific CH to the MS) nodes so as to keep the energy consumption appropriately balanced among all the SNs of the network. Towards the above direction, each SN that receives in our protocol such a forwarded message is forced to send it to its CH through the same path as for its own messages, through Rule 2 (fig. 4). Thus, all the forwarded data will be finally routed to the MS through the CHs of the remaining (till the MS location; in the same direction) clusters, spreading their total load over all the nodes of that clusters.

In other words, through the above protocol all the additional messages that have to be communicated between the CHs, will be evenly distributed through the already constructed increasing-energy paths of the initially formed clusters. This fact, in combination with the fact that the base node-clustering algorithm of [24] (and consequently the corresponding modified algorithm through Rule 1) offers sufficiently balanced energy consumption among all nodes (and correspondingly a-priori energy-balanced paths within each cluster) as one of its key features, guarantees the preservation of high energy efficiency for the whole protocol. Note also that the CHs lying within the MS route (that will be computed later on - see below in section IV) will not have to forward any data to other nodes; since the MS will pass from their locations to collect all the buffered data.

IV.    THE DATA GATHERING PHASE

Once the clustering hierarchy has been established, according to the rules described in the previous section, the MS has then to compute an optimal route for visiting a subset of the elected CHs within the pre-defined time constraint L, and then it can efficiently proceed to periodic data gathering through simple data packet protocols. Apparently, the lower the time constraint L the less the number of CHs the MS will be able to visit. A relevant optimization problem would probably be to find such a time-constrained MS tour, maximizing the number of CHs involved (as in [20]). However this approach is not expected to behave quite well in our case as explained in more details later on (in subsection IV.B). Instead, in our case it's crucial with regard to the nature of the cluster formation algorithm, to include within the constrained tour the CHs that are 'closer to' and 'around' the MS. The latter makes the final solution fit better to the orientation restriction defined by Rule 1 (where 'direction' is defined according to the 'line' from each SN to the MS), and restricts appropriately the number of hops needed for forwarding each message.

### A. Our Basic Data Gathering Protocol

Towards the above direction, we first assume that the MS initially lies at the center of the deployment field, as well as that at the end of the execution of the clustering algorithm, the CHs notify the MS with their exact locations (e.g. by flooding). Let's also denote as C the set of all the elected CHs. We also initially 'sort' all the elements (CHs) of C according to their distance from the initial location of the MS.

that will be visited, in the upper half of C; conversely, if the length constraint is not satisfied, the algorithm continues searching for the right-edge CH that will be visited, in the lower half of C. The whole process is repeated till the optimal right-edge CH position in C is found.

---

**Build MS Tour**

T = 0; first = 0; last = |C|;

left = first; right = last;

c = $\lfloor$ (right - left) / 2 $\rfloor$;

*while*  ((right - left) > 1 and T $\neq$ L)

    Find_TSP_Route (C$_{first}$,C$_c$) ;

    *if*  (T=Found_TSP_Route_Time <= L)

        left = c;

        c = $\lfloor$ (right + c) / 2 $\rfloor$;

    *else*

        right = c ;

        c = $\lfloor$ (left +c) / 2 $\rfloor$;

---

Fig. 6.   Determining the set of visited CHs

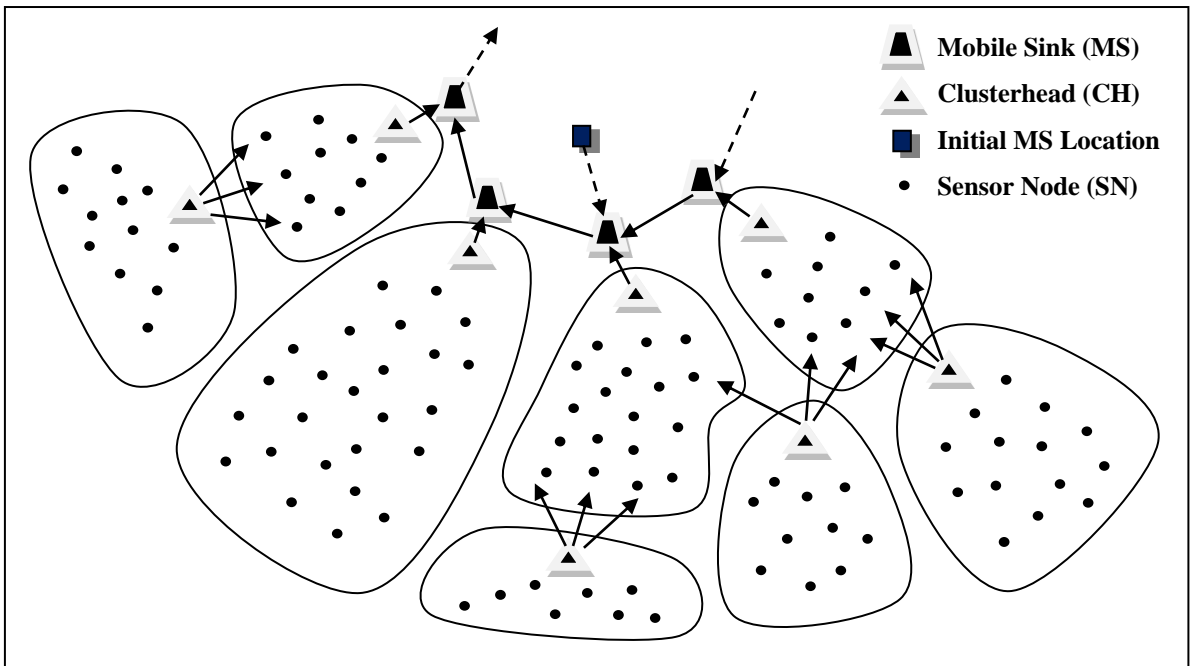

Fig. 5.   Data gathering with the modified clustering (with length constraints)

**Mobile Sink (MS)**
**Clusterhead (CH)**
**Initial MS Location**
**Sensor Node (SN)**

Therefore, the solution can be approximated by finding the most right-positioned CH in C (let's name it $C_e$) for which holds that there is a TSP route (containing the nodes from $C_{first}$ to $C_e$) with tour time <= L. An efficient way to do so (especially for large sizes of C), is following a process similar to the binary-search mechanism (as given in fig. 6). Specifically, the algorithm begins by examining as probable right-edge CH ($C_e$) the CH at the middle of C. If the choice results in a tour that satisfies the constraint L, the algorithm continues searching (in the same way) for the right-edge CH

At the end of execution, *left* holds the value of *c* during the last valid tour, so the final output is the route containing the nodes from $C_{first}$ to $C_{left}$. Function Find_ TSP_Route(), can be efficiently implemented with one of the known TSP heuristics of the literature (e.g. [33]). In this way, a near-optimal tour of maximum time L, around the initial position of the MS, will be computed. Afterwards, the MS can proceed to sequential data gathering rounds over the computed TSP path (like in fig. 5), until a reclustering phase is decided (see below). When it completes each round, it uploads the collected data to the Base Station (BS) and so on.

### B. An Alternative Data Gathering Approach

As mentioned above a natural alternative (followed in many relevant approaches, however not well-suited to our basic MS-oriented protocol) would be to find a corresponding time-constrained MS tour that maximizes the number of CHs involved. This alternative can be easily formulated as a point-to-point orienteering problem (OP [34], with identical scores), and it can also be efficiently implemented with one of the known heuristics [34]. We provide such a complementary solution and we compare it to our basic data gathering protocol described above, in section V.

Although the above approach looks attractive with regard to any MS-based data gathering protocol in WSNs (as it is also shown in our simulation experiments), it doesn't fit properly as part of our total solution, since it's quite possible to conclude to routes that are relatively unstructured and outside the closest possible virtual radius around the initial MS position. As a consequence one or more CHs that are in closer to the initial MS position distance (and within the closest possible virtual radius around that position) are likely to be ignored in such cases. A relevant example is given in Fig. 7. In this example two TSP routes of approximately the same length are presented.

The route with the solid line (including CHs 1,2,3 and 4 – totally four CHs) has been formed adopting our basic protocol, whereas the route with the dashed line (including CHs 4,5,6,7 and 3 – totally five CHs) has been formed adopting the alternative solution (maximizing the number of CHs involved). Obviously the first route (consisting of fewer CHs – four vs five) fits better to our MS-oriented clustering solution, since it forms a route that is very close to a virtual radius around the initial MS position (as opposed to the second route which extends to one side only).

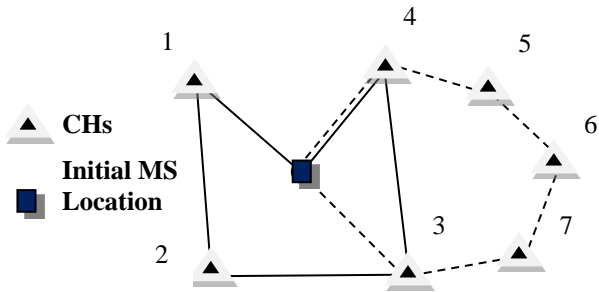

Fig. 7. TSP routes for the two alternatives

### C. Handling the energy holes around the MS

The final step (which relates to another basic problem not taken into account in the solution of [20]) of our solution is to face up effectively the energy holes caused around the CHs involved in the TSP route of the MS. The nodes around these CHs (and the CHs themselves) are the nodes that have to forward the greater loads of data to other nodes (or to the MS), so they are expected to deplete their energy faster than the other SNs. Towards the above direction, we follow a combined solution which consists of,

- applying a suitable global reclustering procedure when needed, according to specific criteria, and

- moving the MS (each time reclustering is decided) to a different initial position, which guarantees that a completely different set of CHs will be visited through the new TSP route.

A decision for reclustering is taken by the MS (similarly to the approach followed in [10]) when it realises that the decrease of the average residual energy of at least one of the clusters of the TSP route, compared to the average residual energy of that cluster at time of last reclustering, is higher than a threshold. In order the MS be able to realise such a situation, each visited CH should simply send to the MS (when it reaches its location) a 'below_threshold' message (along with its sensed data) whenever the reclustering criterion holds for its cluster. Each CH can easily keep track of the average residual energy of its cluster by periodically collecting the necessary information from all its members. Moreover, note that the cluster formation procedure that will be performed during the reclustering will take into account (in order to compute whether each neighbor $v$ lies within the 'direction' between each SN and the MS) the new location of the MS (determined as the result of the second step procedure).

With regard to the detailed execution of the second step above (MS-moving procedure), the new initial location of the MS is specified by first estimating the rectangular area determined by the clusters whose CHs belong to the present TSP route, and then dividing the whole deployment area to such rectangular regions. The MS should then move (each time reclustering is decided) to the next rectangular region (in a suitably predefined manner, e.g. in a cyclic order starting from the center of the filed or in a snake-like order from left to right and up to down). The estimation procedure is repeated each time the MS has to move again, taking into account the present TSP route. In this way, the MS will visit all the regions of the deployment area for specific time intervals, depending on how often reclustering is decided. In each time interval (between two reclustering phases) a different set of CHs will be visited and finally the total data forwarding overhead is expected to be evenly distributed among all the SNs.

In the above, we assume that the MS is capable to upload the gathered data to the BS from any location within the deployment area (e.g. via an internet connection). Also the proposed solution does not take into account probable battery limitations with regard to the available power of the MS, assuming either that the MS is a high-power mobile device, or it has at least enough power for completing the necessary gathering rounds over all the deployment area, or it can be recharged in some way externally in fixed intervals; which are all reasonable assumptions with the current technology.

## V. SIMULATION RESULTS

In the following, we present our extended experimental results taken through simulations with regard to our proposed solution as well as in comparison with the solution of [20], which uses 'distances' (instead of 'residual energy') as the main criterion for data forwarding, and it's one of the most relevant and competent works in the literature. All the simulations have been performed using the Castalia simulator, which is based on the OMNeT++ [35]. We have run experiments for varying number of nodes ($n$=400, 600, 800 and 1000), which are deployed randomly within a square area of side equal to 500m ($500 \times 500m^2$ terrain). The maximum transmission range R of the SNs is equal to 45m and their initial energy is set to 500 Joules. The energy consumption for each transmission depends on the target distance and varies from 29.04mW to 57.42mW (4.3m-45m). The energy consumption for reception and sleep mode is 62mW and 0.016 mW, respectively.

The value of $\theta$ is initially set (for the first set of experiments - subsection V.A) to 70°, which leads to the most satisfactory (close to the best in almost all cases) results for our protocol under the above settings. A detailed analysis with regard to the influence of angle $\theta$ on the performance of our protocol is given in the second set of experiments (subsection V.B). With regard to L, we define four different test values (representing scenarios of corresponding low, medium and high acceptable delays) equal to $0.05T_L$, $0.1T_L$, $0.25T_L$ and $0.5T_L$ respectively, where $T_L$ is defined as the time needed for the MS to visit all the CHs through an optimal TSP route in our protocol (assuming that the MS moves with speed s=1m/s). All the results have been taken as the average out of five independent simulation runs. The above settings are similar to the ones defined in the experiments of [20]. Note also that the energy consumed during the global reclustering procedures required in the proposed data gathering scheme (subsection IV.C) as well as for any other control message transmission, has been included in all the corresponding measurements.

### A. Basic experimental results (# of SNs, L value)

Our basic experimental results are summarized in figures 8-11. First, in fig. 8, the network lifetime achieved by our protocol in case of no delay constraints (in this case we simply execute the initial clustering algorithm of section II, and the MS is scheduled to visit all the elected CHs) is given in comparison with the network lifetime achieved by the protocol of [20] in the same case. As it can be seen the network lifetime achieved by our protocol is clearly higher for all the numbers of SNs within the terrain. The corresponding differences are over 10% in all the test cases (from 11% to 19,5%, approximately). Moreover, in our protocol the network lifetime remains almost the same as the number of sensors increase. This happens due to the stable behavior of the initial clustering structure, which keeps both the average energy consumption almost constant, as well as the variance of the residual energy very low and almost constant too.

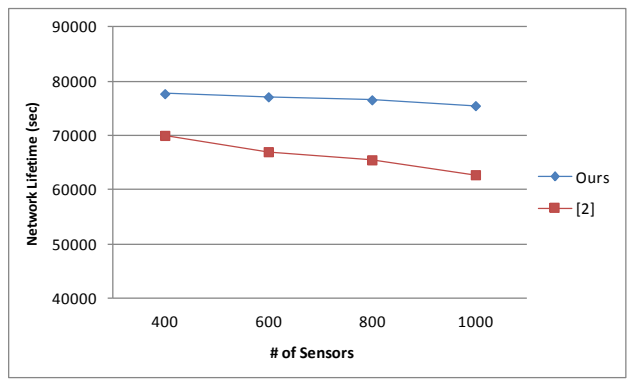

Fig. 8.   Network lifetime with no constraints

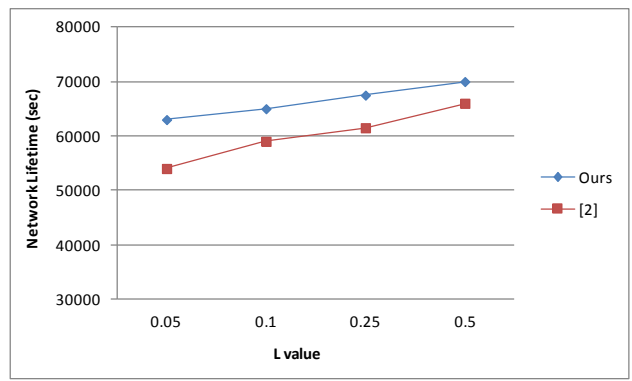

Fig. 9.   Network lifetime vs *L* for *n*=800

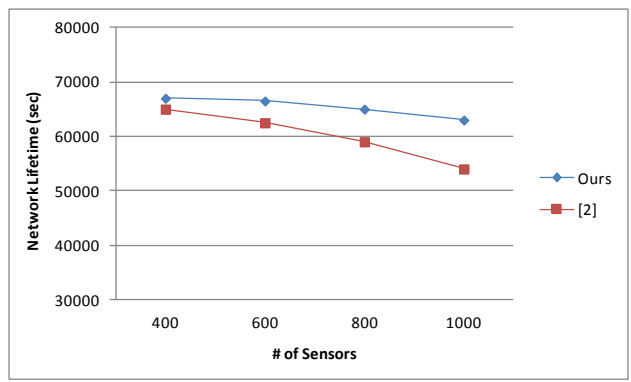

Fig. 10.   Network lifetime vs *n* for *L*=0.1

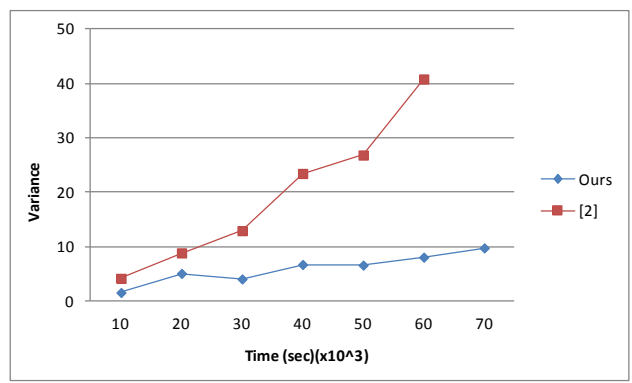

Fig. 11.   Variance for *n*=800, *L*=0.1

Proceeding to the experiments with specific delay constraints (where we use the modified clustering algorithm given in section III and the basic data gathering protocol given in subsection IV.A), we present in fig. 9 the network lifetime achieved by the two protocols, for varying values of L (delay constraint) and for constant number of nodes (equal to 800). As it can be easily noticed for medium and small values of L the difference in favor of our protocol is clear (up to 15,5%), whereas for larger values of L it decreases a lot (down to 6%). Also, the network lifetime of our protocol increases with the increase of L due to the fact that as L increases the MS visits more and more CHs, thus reducing the amount of messages (communication load) that have to be forwarded through other clusters.

Furthermore, in fig. 10 the network lifetime for the two protocols is presented, for varying values of nodes and for constant value of L (equal to $0.1T_L$). As it can be easily noticed here too, for medium and large numbers of nodes the difference in favor of our protocol is significant (up to 17%), whereas for smaller number of nodes the behavior of the two protocols is almost the same. Also the network lifetime of our protocol appears to decrease a little with the increase of the number of SNs. This happens because as the number of SNs increases, the CHs and cluster members that have to forward messages of other clusters too (except their own messages), are overloaded in a quite more intensive way; so they normally deplete their energy in a less controllable way, despite of the recovering measures taken - global reclustering and MS initial position movement.

Finally, in fig. 11 the variance of the residual energy is presented for the two protocols, during the network lifetime, and for constant number of nodes and value of L (800 and 0.1 respectively, which represent a medium-to-large instance). As it can be seen, the variance for our protocol is quite low, during the whole network lifetime, and much lower than in the protocol of [20]. The latter means that in our protocol the energy of all nodes depletes in a much more uniform way than in the other protocol. This naturally explains the significant differences in the network lifetime presented in figures 8-10.

As a general conclusion, our data gathering protocol is shown to behave considerably better in all the testing cases, except the case of large values of L and small number of nodes - density - where the two protocols have similar behavior. The energy depletion of all nodes in our protocol is sufficiently uniform; as opposed to the protocol of [20] in most cases. This naturally results in significant increase of the network lifetime in almost all testing cases. In [20] the fact (a) that the energy holes caused around the CHs lying within the MS route are not handled, in combination with the fact (b) that the corresponding clustering algorithm itself (along with the data forwarding protocol) does not take in account the residual energy of the SNs, naturally leads to non-uniform energy depletion of the SNs as the test case becomes more intensive. As a result the protocol of [20] behaves well for smaller number of nodes and higher values of L, whereas, for larger instances and small delays its performance decrease significantly; in these cases the node-trees formed around each CH in [20] are getting larger and larger pointing out the above referred disadvantages of the corresponding solution.

### B. Measurements for different values of $\theta$

As discussed in section III, the exact value of angle $\theta$ is a crucial factor that influences significantly the performance of the proposed protocol. Specifically, the higher the value of $\theta$ the broader the area in front of each CH in which the total load of forwarded data will be distributed (and the better the balancing of the energy consumption achieved). However also, the higher the value of $\theta$ the higher the probability of concluding to longer paths for some of the communicated messages, thus loosing the advantage gained from the ideally balanced energy consumption that lies in the heart of our algorithm. Moreover, one can easily realize that the 'best' value for angle $\theta$ depends also on the value of L. Specifically, the larger the value of L the larger the expected 'best' value for $\theta$, since the MS will normally visit CHs that are spread on a larger virtual radius (with respect to the initial MS point).

In other words, for different values of L different 'best' values for angle $\theta$ are normally expected. We've performed relevant experiments to explore the performance of our protocol for different values of $\theta$ (and keeping constant in each case either the number of nodes or the value of L). The corresponding measurements are presented in fig. 12 and 13.

In fig. 12 the network lifetime is given for a wide range of values of $\theta$ (45°, 60°, 67.5°, 75° and 90°) and for all the different values of L (0.05$T_L$, 0.1$T_L$, 0.25$T_L$ and 0.5$T_L$), whereas the number of nodes is kept constant (n=800). As it can be seen, the network lifetime increases with the increase of L until a maximum is reached (the best value of $\theta$ for that case). Then it clearly decreases as it approaches to 90°, which is an angle value that leads to an extremely spread area in front of each node/CH for data propagation, thus concluding to significantly longer (in total number of hops) forwarding paths.

More concretely, the best value of $\theta$ for L=0.05$T_L$ is around 65°, whereas for the other values of L (0.1$T_L$, 0.25$T_L$, 0.5$T_L$) the exact best value of $\theta$ is around 70°, 71° and 75° respectively. The corresponding exact maximum values computed for angle $\theta$ through the complete set of our simulation experiments were 65.3°, 70.2°, 71.1° and 75.7° respectively. As it was expected, as the value of L increases the 'best' value for angle $\theta$ increases too. However the corresponding range is quite closed (from 65° to 75° approximately); i.e. it doesn't vary proportionally to the value of L or any other factor. Moreover, it must be noted that the desired time constraint (value of L) as well as the other settings of the network are normally a-priori known in a realistic application, so the 'best' value of $\theta$ can easily be determined or at least approximated.

Furthermore, in fig. 13 the network lifetime is given for the same range of values of $\theta$ (45°, 60°, 67.5°, 75° and 90°) and for all the different numbers of SNs (400, 600, 800 and 1000), whereas the value of L is kept constant (L=0.1$T_L$). As it can be seen, the best value for angle $\theta$ is approximately the same (around 70°) in all cases. This happens because the number of clusters (and CHs) in our clustering structure is not affected significantly (it remains almost the same) by the increase of the number of nodes (density) within the same deployment area; so the MS tour is expected to be quite similar in all cases.

### C. Comparing the two TSP route alternatives

Finally, we compare our basic protocol to the other possible alternative discussed in subsection IV.B with respect to data gathering, i.e. finding the time-constrained MS tour that maximizes the number of CHs involved. The corresponding measurements are presented in fig. 11 and 12. In fig. 11 the network lifetime is given for both the two alternatives and varying number of SNs, whereas in fig. 12 the network lifetime is given for varying value of L. In the first case (fig. 11) the value of L is kept constant (L=0.1$T_L$), whereas in the second case (fig. 12) we keep constant the number of SNs (n=800). In both cases the value of $\theta$ is set to 70°.

As it can be seen in both figures the network lifetime achieved by our basic protocol (alter-1) is clearly higher in almost all cases, either for varying numbers of SNs or for varying L. More concretely, for small number of SNs and small tours (small values of L) the network lifetime achieved by the two alternatives is almost the same (slightly better for alter-1). This happens because in these cases the two alternatives normally lead to very similar TSP routes. On the other hand for larger number of SNs and larger tours the difference is much more clear, raising up to almost 7% for 1000 nodes and almost 10% for L=0.5T$_L$. The reason for the above differences lies on the fact that in these cases trying to maximize the number of CHs involved, we may easily conclude (as also explained in subsection IV.B) to relatively unstructured routes, lying outside the closest possible virtual radius around the initial MS position, and naturally ignoring one or more CHs that are in closer to that position distance.

## VI. CONCLUSION

A residual energy based data gathering solution for WSNs with delay constraints is presented throughout this paper. The heart of our solution is an energy-efficient multi-hop clustering algorithm appropriately modified by taking into account the orientation of the SNs with respect to the location of the MS. The energy balanced clusters formed due to the nature of the clustering procedure, along with the followed data forwarding algorithm, and the applied reclustering and MS movement procedures, guarantee the sufficiently balanced energy consumption among all nodes and the preservation of high energy efficiency for the whole protocol. Based on extended simulation experiments our protocol is shown to be highly stable and efficient, whereas also, it achieves considerably better network lifetime than other competent approaches in the literature (like the one presented in [20]). The efficient use of multiple mobile sinks (or mobile relay nodes) in the proposed data gathering approach is of high priority in our future work. We plan to design and evaluate such generalized solutions in order to provide an efficient alternative for very large WSN applications. We also plan to extend the proposed MS-oriented clustering algorithm for heterogeneous WSN environments with various limitations. Finally, several other practical alternatives with regard to the MS-moving procedure (to a different initial position each time reclustering is decided) should be implemented and evaluated, targeting to completely eliminate the overhead caused by the energy holes around the MS.

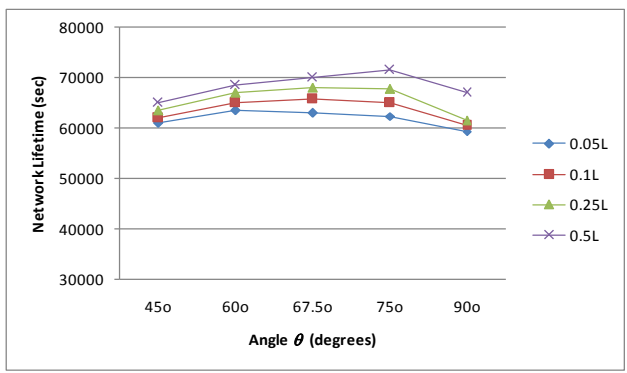

Fig. 12. Network lifetime for varying θ and L (n=800)

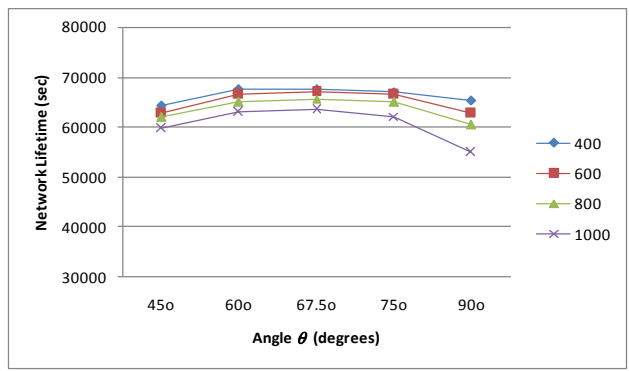

Fig. 13. Network lifetime for varying θ and n (L=0.01)

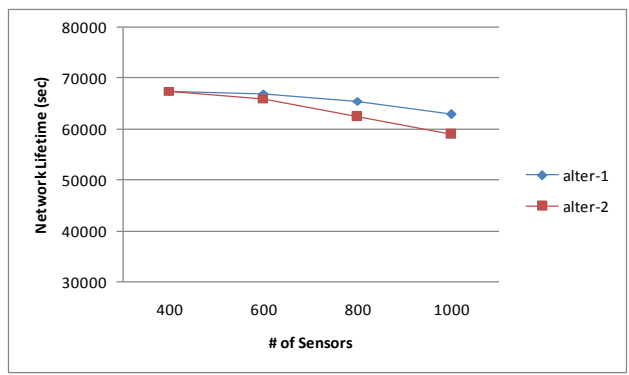

Fig. 14. The two alternatives for varying n (L=0.01)

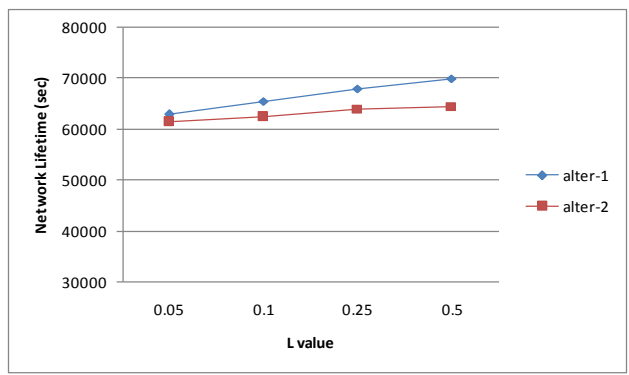

Fig. 15. The two alternatives for varying L (n=800)